\documentclass[10pt,aps,twocolumn,superscriptaddress]{revtex4-1}
\usepackage{amsmath,amssymb,mathdots, graphicx, color}
\usepackage[margin=1in]{geometry}
\newcommand{\ket}[1]{\rvert#1\rangle}
\newcommand{\bra}[1]{\langle #1\rvert}
\newcommand{\braket}[2]{\langle #1\rvert#2\rangle}

\newmuskip\pFqskip
\mathchardef\pFcomma=\mathcode`, % keep a copy of the comma

\newcommand*\pFq[5]{%
  \begingroup
  \begingroup\lccode`~=`,
    \lowercase{\endgroup\def~}{\pFcomma\mkern\pFqskip}%
  \mathcode`,=\string"8000
  {}_{#1}F_{#2}\biggl[\genfrac..{0pt}{}{#3}{#4};#5\biggr]%
  \endgroup
}
\begin{document}
\title{Analytic next-to-nearest neighbour XX models with perfect state transfer and fractional revival}
\author{Matthias Christandl}
\affiliation{Department of Mathematical Sciences, University of Copenhagen, Universitetsparken 5, 2100 Copenhagen, Denmark}
\author{Luc Vinet}
\affiliation{Centre de Recherches Math\'ematiques, Universit\'e de Montr\'eal, C.P. 6128, Succ. Centre-ville, Montr\'eal, QC, Canada, H3C 3J7}
\author{Alexei Zhedanov}
\thanks{On leave of absence from Donetsk Institute for Physics and Technology, Donetsk 83114, Ukraine}
\affiliation{Department of Applied Mathematics and Physics, Graduate School of Informatics, Kyoto University, Kyoto 606-8501, Japan}
%\date{}

\begin{abstract}
Certain non-uniformly coupled spin chains can exhibit perfect transfer of quantum states from end to end. Motivated by recent experimental implementations, we extend the simplest such chain to next-to-nearest neighbour (NNN) couplings. It is shown analytically that perfect state transfer can be maintained under the extension and that end-to-end entanglement generation (fractional revival) can occur. 
%It is surmised that this offers a relevant model for the implementation of quantum wires by waveguide arrays. It is moreover predicted, in distinction to what the dynamics of the model with only nearest neighbour interactions shows, that optical lattices should generate maximally entangled states via FR.
\end{abstract}

\maketitle

\section{Introduction}
\subsection{Background}
  The transfer of quantum states between remote sites is an important issue in the development of quantum technologies.  Theoretical studies have shown that over short distances spin chains could achieve this task with a minimum of control operations (see \cite{1} and the recent review  \cite{28}). Indeed, with properly engineered couplings, the intrinsic dynamics of the chain can realize the transfer with unit probability, in which case one speaks of perfect state transfer (PST) \cite{3, 2, 4}. Another phenomenon, known as fractional revival (FR) \cite{5}, can also be observed in spin chains \cite{5a,5b,2,6,5c,7} and exploited for quantum communication purposes. In this case, an excitation on one end of the chain splits into a superposition of an excitation at each end, resulting in entanglement between the ends of the chain.

  The class of spin chain Hamiltonians that has been principally examined in this connection is that of type XX with nearest-neighbour (NN) couplings,
\begin{align}
   H = \frac{1}{2}\sum_{n=0}^{N-1} J_{l}(\sigma_n^x\sigma_{l+1}^x+\sigma_n^y\sigma_{n+1}^y) + \frac{1}{2}\sum_{n=0}^N B_n(\sigma_n^z+1),
  \end{align}
  where the $\sigma^{x / y / z}_l$ denote the corresponding Pauli matrix acting on site $l$.
  Owing to rotational symmetry about one axis, the total spin projection is conserved in these models and the space of states decomposes into a sum of subspaces labelled by the number of excitations. PST is then determined by the one-excitation dynamics. Restricting to this dynamics, $H$ takes the form of a tridiagonal matrix given by
 \begin{align} 
  \begin{pmatrix} \label{2}
  B_0	&	J_1	&		&	&
  \\
  J_1	&	B_1	&	J_2	&	&
  \\
	  &	J_2	&	B_2	& \ddots &
  \\
	  & 		& \ddots		& 	\ddots & J_{N}
  \\
  & & &J_{N}&B_{N}
  \end{pmatrix}.
    \end{align}
 Families of orthogonal polynomials that have the matrix entries as their recurrence coefficients provide the exact eigenfunctions \cite{8, 4}. Consequently, spin chains with perfect state transfer have been investigated systematically and many types have been found. The simplest and thus paradigmatic model in this context is associated to the Krawtchouk polynomials. The couplings in this case follow an elementary parabolic distribution \begin{align}\label{1, 46}
 J_n = a_n \equiv \frac{1}{2}\sqrt{n(N-n+1)}
\end{align}
and there are no Zeeman terms, $B_n=0$. The Krawtchouk chain does not exhibit fractional revival, but models that do have been found and indeed been analysed systematically \cite{5c,7,5d}.

\subsection{Recent experiments}
Perfect state transfer has been demonstrated recently in different technological platforms. Perfect state transfer and the stronger mirror inversion have been observed using NMR techniques \cite{9}. Since the time evolution of a single excitation is mathematically equivalent to an excitation in a lattice of oscillators in the tight binding formalism with nearest-neighbour approximation (only when multimodes are considered differences between fermions and bosons appear), photonic lattice implementations of perfect state transfer have been considered \cite{10,11,12,27}. Indeed, it has been demonstrated experimentally that quantum states can be transported with high fidelity through arrays of evanescently coupled waveguide elements \cite{13,14,15}. The present paper is largely motivated by these advances. Since the evanescent couplings of the waveguides in an optical array depend on the separation between the components, interactions always extend beyond the nearest neighbours and a better approximate description would involve next-to-nearest couplings. It could also be that manufacturing or setting imprecisions would have the array in a slight zigzag pattern in which case the NNN couplings can become even more significant relative to the NN couplings \cite{16}. The same could happen of course if the quantum network is bent \cite{17}.

These considerations call for an examination of PST in models with NNN interactions in addition to the NN ones. Such a study would provide a framework to better estimate the errors made when the NNN couplings are neglected, it could inform the situations when these couplings are really required and finally, it could unveil new dynamical possibilities when NNN interactions are present. The question of PST beyond NN couplings has been addressed in \cite{18} where it has been shown that inverse spectral problem methods can be applied to obtain Hamiltonians with interactions beyond the nearest neighbours that exhibit PST. See also \cite{19} in this connection. This work, however, is numerical and, to our knowledge, no such analytic models have been identified so far. 

\subsection{Our contribution}
  The purpose of this paper is precisely to fill that lacuna and to offer a simple analytic model with PST and also FR that naturally extends the Krawtchouk model. The nearest-neighbour couplings $J_n^{(1)} $ of this new model remain the same as those of the Krawtchouk XX spin chains with a scaling factor $\beta$
  $$J_n^{(1)} = J_n = \beta a_n,$$ 
  the next-to-nearest neighbour couplings $J_n^{(2)}$ involve an additional parameter $\alpha$ and non-uniform magnetic fields are present. 
  \begin{align}
J_n^{(2)} &= \alpha a_{n-1}a_n \qquad B_n = \alpha(a_n^2+a_{n+1}^2).
\end{align}
  
  PST will be shown to happen when $\tfrac{\alpha}{\beta}$ is rational. For some of these values, it might be necessary to have an odd or even number of sites. If $\tfrac{\alpha}{\beta}$ is an integer, it will be seen that the minimal time for PST in the NNN chains is the same as the one with the NN couplings. 
 %A discussion will be offered as to how this model with NN and NNN couplings can be applied to the description of light propagation in arrays of evanescently coupled single-mode waveguides. A conclusion is that in some circumstances, depending on the specifications of the photonic lattice and on the experimental conditions, better theoretical PST fidelity will be obtained at a time different than the minimal time suggested by the NN model. 
 It will further be seen that in many instances, the model with the NNN couplings will exhibit FR. This is remarkable because this phenomenon does not take place in the Krawtchouk chain with only nearest-neighbour couplings and might be of experimental relevance.
 %It implies that the device can be used to generate maximally entangled states. This hence suggests that FR should happen in optical lattices. We believe that an experimental validation of this prediction would be of significant interest. 
  
  \subsection{Paper structure}
  The paper is structured as follows. In Section \ref{sec:prelim} we introduce the basics of the non-uniform XX chain and review in more detail mirror symmetry in the nearest neighbour situation. We also introduce the special class of next-to-nearest neighbour chains we will consider in this paper. In Section \ref{sec:PST}, we study general conditions for perfect state transfer in XX chains and discuss in detail the well-known Krawtchouk case for nearest neighbour couplings. We then characterise as one main result of this study perfect state transfer in the next-to-nearest neighbour extension of the Krawtchouk case. In Section \ref{sec:FR} we turn our attention to fractional revival and give also here an exact characterisation of our Krawtchouk extension. In Section \ref{sec:conclusion} we offer a conclusion and an outlook.

\section{Preliminaries} \label{sec:prelim}
\subsection{Spin chains of XX type}
  We turn first to the spin chain context and consider the following Hamiltonians of type XX on $(\mathbb{C}^2)^{\otimes(N+1)}$ where each of the $(N+1)$ spins interacts with its $M$ neighbours on the left and $M$ neighbours on the right:
  \begin{align*}
   H^{(M)} = \frac{1}{2}\sum_{l=0}^{N-1}\sum_{k=1}^M J^{(k)}_{l+k}(\sigma_l^x\sigma_{l+k}^x+\sigma_l^y\sigma_{l+k}^y)\\ + \frac{1}{2}\sum_{l=0}^N B_l(\sigma_l^z+1).
  \end{align*}
  The constant $J^{(k)}_{l+k}$ is the coupling strength between the site $l$ and $l+k$ and is taken to be non-negative; the constant $B_l$ is the value of the magnetic field at the site $l$. 
  %The symbols $\sigma_l^x, \sigma_l^y, \sigma_l^z$ denote the Pauli matrices with $l$ indexing the $\mathbb{C}^2$ factor on which they act. 

  Since, the $z$-component of the total spin is conserved, 
  \begin{align}
   [H^{(M)},\frac{1}{2}\sum_{l=0}^N (\sigma_l^z+1)]=0,
  \end{align}
  the eigenstates of $H^{(M)}$ belong to subspaces labelled by the number of spins over the chain that are up, i.e. that are eigenstates of $\sigma^z$ with eigenvalue $+1$. The state of $H^{(M)}$ with all the spins down has energy zero. It will suffice for our purposes to consider states with only one spin up. A natural basis for that subspace (equivalent to $\mathbb{C}^{(N+1)}$) is given by the vectors
  \begin{align} \label{3}
   \ket{n} = (0,0,\dots,1,\dots,0)^T
  \end{align}
  $n=0,\dots,N$, with the only $1$ in the n\textsuperscript{th} position corresponding to the only spin up at the n\textsuperscript{th} site. The action of $H^{(M)}$ on those states is given by
  \begin{align} \begin{aligned} \label{4}
   H^{(M)}&\ket{n} = \\[0.2em]&J^{(M)}_{n+M}\ket{n+M} + J^{(M-1)}_{n+M-1}\ket{n+M-1} + \dots \\[0.2em]&+ J^{(1)}_{n+1}\ket{n+1} + B_n\ket{n}+ J^{(1)}_{n}\ket{n-1}\\&+\dots + J^{(M)}_{n}\ket{n-M}. %\notag
  \end{aligned}\end{align}
  The conditions $J_l^{(k)}=0$ for $l\leq k-1$ and $J^{(k)}_{N+1} =0$ are assumed. 

  In the following we shall consider models whose non-nearest neighbour couplings are constructed from the interaction strengths between the nearest neighbours. We shall hence make much use of the description of the single excitation dynamics when only nearest neighbours are coupled; its essentials are summarized in the next subsection.
 
\subsection{Nearest-neighbour interactions}

When $M=1$, there are only nearest-neighbour interactions. Let us denote by $J$ the restriction of $H^{(1)}$ to the single-excitation eigensubspace. It is readily seen that \eqref{4} specializes to
\begin{align} \label{5}
 H^{(1)}\ket{n} \equiv J\ket{n} = J_{n+1}\ket{n+1} + B_n\ket{n} + J_n\ket{n-1}\phantom{\sum}
\end{align}
 where we have replaced $J_n^{(1)}$ by $J_n$ for simplicity. We thus see that in the occupation basis, $J$ can be represented by the $(N+1)\times(N+1)$ tridiagonal matrix \eqref{2} given above.
%\begin{align} \label{6}
%  J=
%  \begin{pmatrix}
 % B_0	&	J_1	&		&	&
 % \\
 % J_1	&	B_1	&	J_2	&	&
 % \\
%	  &	J_2	&	B_2	& \ddots &
%  \\
%	  & 		& \ddots		& 	\ddots & J_{N}
%  \\
%  & & &J_{N}&B_{N}
 % \end{pmatrix}.
%\end{align}
Introduce the normalized eigenvectors $\ket{x_s}$ of $J$ :
\begin{align}
 J\ket{x_s} = x_s\ket{x_s}, \qquad s=0,1,\dots,N.
\end{align}
The eigenvalues $x_s$ are real and non-degenerate: $x_s\neq x_t$ if $s\neq t$, for positive couplings. Let us expand the eigenstates $\ket{x_s}$ in terms of the basis vectors $\ket{n}$ :
\begin{align} \label{8}
 \ket{x_s} = \sum_{n=0}^N W_{sn}\ket{n}
\end{align}
and write the expansion coefficients as follows :
\begin{align}
 W_{sn} = W_{s0} \chi_n(x_s)
\end{align}
with $\chi_0(x_s) =1$ by construction. Set $\sqrt{w_s}=W_{s0}$. Relation \eqref{8} becomes
\begin{align} \label{10}
 \ket{x_s} = \sum_{n=0}^N \sqrt{w_s}\chi_n(x_s)\ket{n}.
\end{align}
It follows from \eqref{5} that the coefficients $\chi_n(x)$ obey the 3-term recurrence relation 
\begin{align} \label{11}
 J_{n+1}\chi_{n+1}(x) + B_n\chi_{n}(x) + J_n\chi_{n-1}(x) = x\chi_n(x)\phantom{\sum}
\end{align}
and are hence orthogonal polynomials with the initial condition $\chi_{-1} =0$ added.
Since the eigenbasis and the occupation basis are both orthonormal and given that the expansion coefficients are real, these elements form an orthogonal matrix and provide the inverse relations
\begin{align} \label{12}
 \ket{n} = \sum_{s=0}^N \sqrt{w_s}\chi_n(x_s)\ket{x_s}.
\end{align}
Now the fact that $\braket{n}{m}=\delta_{nm}$ implies that
\begin{align} \label{13}
 \sum_{s=0}^N w_s\chi_n(x_s)\chi_m(x_s) = \delta_{nm}
\end{align}
showing that the polynomials $\chi_n(x)$ are orthogonal over the finite set of eigenvalues $x_s$ of $J$ with weight $w_s$.
From the eigenvalues $x_s$, $s=0,1,\dots,N$ one can form the characteristic polynomial $P_{N+1}(x)$ of degree $N+1$ :
\begin{align}
 P_{N+1}(x)= (x-x_0)(x-x_1)\dots(x-x_N)
\end{align}
which is obviously orthogonal to all $\chi_n(x)$ $n=0,1,\dots,N$ according to \eqref{13}. It is known from the theory of orthogonal polynomials \cite{24} that the discrete weights $w_s$ can be expressed as
\begin{align} \label{15}
 w_s = \frac{\sqrt{h_N}}{\chi_N(x_s)P'_{N+1}(x_s)} \qquad s=0,1,\dots,N,
\end{align}
where 
\begin{align}
 \sqrt{h_N} = J_1J_2\dots J_N
\end{align}
and where $P'_{N+1}(x)$ stands for the derivative of $P_{N+1}(x)$. If one takes the eigenvalues in increasing order,
\begin{align}
 x_0<x_1<\dots<x_N,
\end{align}
it is easy to see that
\begin{align} \label{18}
 P'_{N+1}(x_s) = (-1)^{N+s}|P'_{N+1}(x_s)|.
\end{align}
The class of mirror-symmetric couplings and field strengths satisfying
\begin{align} \label{19}
 J_n = J_{N+1-n}, \qquad B_n = B_{N-n}
\end{align}
is central in PST considerations. (See for instance \cite {2,4,8}.)  In terms of the matrix $J$ given in \eqref{2}, the conditions \eqref{19} amount to the requirement that $J$ be reflection invariant with respect to its main anti-diagonal and thus satisfy $RJ=JR$ with
\begin{align} \label{20}
  R = 
  \begin{pmatrix}
    0& \dots & 0 & 1  \\
   0&\dots  & 1& 0  \\
    & \iddots & &  \\
    1&0 &\dots &0  \\
  \end{pmatrix}
\end{align}
the matrix transcription of the reflection defined by $R\ket{n} = \ket{N-n}$. Observe that $R^2=1$. When $J$ and $R$ commute, they can be diagonalized simultaneously and since the eigenstates $\ket{x_s}$ are non-degenerate we have then 
\begin{align} \label{21}
 R\ket{x_s} = \epsilon_s\ket{x_s}
\end{align}
with $\epsilon_s = \pm 1$. Moreover, from \eqref{10} we obtain 
\begin{align} \label{22}
 R\ket{x_s} &= \sum_{n=0}^N \sqrt{w_s}\chi_n(x_s)\ket{N-n} \notag\\&= \sum_{n=0}^N \sqrt{w_s}\chi_{N-n}(x_s)\ket{n}.
\end{align}
Eqs \eqref{21} and \eqref{22} thus yield 
\begin{align} \label{23}
 \chi_{N-n}(x_s) = \epsilon_s \chi_n(x_s).
\end{align}
It can also be shown \cite{8} that in the presence of mirror symmetry the weights are given by 
\begin{align} \label{24}
 w_s = \frac{\sqrt{h_N}}{|P'_{N+1}(x_s)|}.
\end{align}
Comparing \eqref{24} with the general formula \eqref{13} and using \eqref{18} allows to show that
\begin{align} \label{25}
 \chi_N(x_s) = (-1)^{N+s}
\end{align}
is equivalent to the condition that the matrix $J$ be mirror-symmetric. Setting $n=0$ in \eqref{23}, one further concludes that 
\begin{align}
 \epsilon_s = (-1)^{N+s}
\end{align}
and then obtains from \eqref{23} the following property of orthogonal polynomials associated to mirror-symmetric Jacobi matrices : 
\begin{align}\label{27}
 \chi_{N-n}(x_s) = (-1)^{N+s} \chi_N(x_s).
\end{align}

\subsection{From nearest-neighbor to next-to-nearest-neigbor couplings}
As seen from eq.\eqref{4}, the couplings and magnetic field strengths of the Hamiltonians $H^{(M)}$ are in fact determined from the restriction of $H^{(M)}$ to the one-excitation subspace. In the following, we shall consider a special class of $H^{(M)}$ where the non-nearest neighbour couplings are prescribed from those between the nearest neighbours. We shall define these special Hamiltonians by demanding that their restriction to states with only one spin up takes the form
\begin{align} \label{28} 
H^{(M)}\ket{n} = Q_M(\bar{J})\ket{n}                                                                                                                                                                                                                                                                                                                                                                                                                                                                                                                                                                                                                                                                                                                                                                                                                                                                                                                                                                                  \end{align}
where $Q_M$ is a polynomial of degree $M<N$ in a fiducial tridiagonal matrix $\bar{J}$ of the same form as \eqref{2} with entries that can be viewed as basic nearest neighbour couplings and field strengths. From the action of $\bar{J}$ on $\ket{n}$ given by \eqref{5} with $J_n$ and $B_n$ replaced by $\bar{J_n}$ and $\bar{B}_n$ respectively, it is straightforward to compute $Q_M(\bar{J})\ket{n}$ and to identify the couplings $J_l^{(k)}$ of $H^{(M)}$. This will provide families of Hamiltonians where the spin at site $n$ interacts with its $M$ neighbours on the left and on the right; we shall henceforth limit our considerations to this special class of systems with non-nearest neighbour interactions. 

Note that the formalism developed above applies generally to tridiagonal matrices and hence to $\bar{J}$. In the following, we shall wish to distinguish between $\bar{J}$ and $J^{(1)}=J$. Typically we shall have $J=\beta \bar{J}$ with $\beta$ a constant. For clarity, when using the results of Section \ref{sec:prelim} as they pertain to $\bar{J}$, we shall denote by $\bar{x}_s$ the eigenvalues of $\bar{J}$, by $\bar{\chi}_n$ the orthonormalized polynomials associated to $\bar{J}$, etc.

\section{Perfect state transfer}\label{sec:PST}

\subsection{General considerations}

Let us now explore the conditions for the transfer with probability one, after time $T$, of a spin up from one end of the chain to the other, when the dynamics is governed by the special Hamiltonians $H^{(M)}$ defined by \eqref{28}. This PST will be realized if
\begin{align} \label{29}
 e^{-iTH^{(M)}}\ket{0} = e^{i\phi}\ket{N}
\end{align}
where $\phi$ is some phase. With the help of the expansion \eqref{12} and with a bar over the attributes of $\bar{J}$, we see that condition \eqref{29} amounts to
\begin{align} \label{30}
 e^{-i\phi}\exp[-iTQ_M(\bar{x}_s)] = \bar{\chi}_N(\bar{x}_s)
\end{align}
which entails restrictions on the spectrum of $\bar{J}$. Since the rhs of \eqref{30} only takes real values, we must have
\begin{align} \label{31}
 \bar{\chi}_N(\bar{x}_s) = \pm1 \qquad s=0,1,\dots,N.
\end{align}
Let $\{ p_k \}$ be a sequence of orthogonal polynomials; for $y_1<y_2<\dots<y_n$ and $z_1<z_2<\dots<z_{n-1}$, the zeros of $p_n$ and $p_{n-1}$ respectively, it is well known \cite{20} that one has 
\begin{align} \label{32}
 y_1<z_1<y_2<z_2<\dots<z_{n-1}<y_n, 
\end{align}
a property called the interlacing of the zeros of $p_n$ and $p_{n-1}$. Applying this result to $\bar{P}_{N+1}$ and $\bar{\chi}_N$, one sees that a zero of $\bar{\chi}_N$ must be encountered between each of the pairs $(\bar{x}_{s-1},\bar{x}_{s})$, $s=1,2,\dots,N$ and hence, in view of \eqref{31}, the sign of $\bar{\chi}_N$ must alternate at successive eigenvalues $\bar{x}_s$. Once this is established, it follows from \eqref{15} and \eqref{18} that we must have $\bar{\chi}_N(\bar{x}_s) = (-1)^{N+s}$ for the weights $\bar{w}_s$ to be positive. This is precisely eq.\eqref{25} which, as noted in Section \ref{sec:prelim}, is tantamount to $\bar{J}$ being mirror-symmetric. We thus conclude that for the special class of chains with non-nearest couplings we are considering, a necessary condition for PST is that the underlying Jacobi matrix satisfies $\bar{J}R=R\bar{J}$. This of course implies that the one-excitation restriction of $Q_M(\bar{J})$ of $H^{(M)}$ itself commutes with $R$ and that the couplings satisfy
\begin{align} \label{33}
 J_n^{(k)} &= J_{N-n+k}^{(k)} &&k=1,\dots,M. \\
 B_n &= B_{N-n}  &&k\le n\le N. \notag
\end{align}
Obviously when $N=1$, this amounts to \eqref{19} which is known to be one of the necessary conditons for PST when only nearest neighbour interactions are present. Given the requirement that \eqref{25} be satisfied, condition \eqref{30} remains and becomes 
\begin{align} \label{34}
 e^{-i\phi}\exp[-iTQ_M(\bar{x}_s)] = (-1)^{N+s}.
\end{align}
This translates into
\begin{align} \label{35}
 TQ_M(\bar{x}_s) = -\phi + \pi(N+s+2L_s) \quad s=0,\dots,N
\end{align}
where $L_s$ are arbitrary integers that may depend on $s$. Eq \eqref{35} leads to restrictions on the spectrum of $\bar{J}$ and the coefficients of $Q_M$; together with the mirror symmetry of $\bar{J}$, it provides the necessary and sufficient conditions for PST in the case under study. In fact, these requirements ensure the full mirror inversion of the one-excitation states after time $T$. To see this, consider the matrix elements $\bra{k}e^{-iTH^{(M)}}\ket{l}$ and use the expansion \eqref{12} over the eigenstates of $\bar{J}$, one has
\begin{align} \label{36}
 \bra{k}e^{-iTH^{(M)}}\ket{l} = \sum_{s=0}^N e^{-iTQ_M(\bar{x}_s)}\bar{w}_s\bar{\chi}_k(\bar{x}_s)\bar{\chi}_{l}(\bar{x}_s).
\end{align}
Using \eqref{34}, \eqref{27} and the orthogonality relation \eqref{13} (for the polynomials $\bar{\chi}_{n}$), one obtains
\begin{align} \label{37}
 \bra{k}e^{-iTH^{(M)}}\ket{l} &= e^{i\phi}\sum_{s=0}^N \bar{w}_s\bar{\chi}_k(\bar{x}_s)\bar{\chi}_{N-l}(\bar{x}_s) \notag\\&= e^{i\phi}\delta_{k,N-l}
\end{align}
which implies as announced that
\begin{align} \label{38}
 e^{-iTQ_M(\bar{J})} = e^{i\phi}R.
\end{align}

We shall now further restrict the set of models that we will consider by fixing the matrix $\bar{J}$ and examining PST first for nearest-neigbour and then when next to nearest neighbour couplings are added, that is when $M=2$.

\subsection{The Krawtchouk chain}
The simplest possible spectrum that can be posited for a mirror-symmetric Jacobi matrix is the linear one where
\begin{align} \label{39}
 \bar{x}_s = s-\frac{N}{2} \qquad s=0,1,\dots,N.
\end{align}
The methods of inverse spectral problems can be used to obtain the corresponding $\bar{J}$ \cite{4}. This matrix can however be simply identified through the following observations. (Among the many references on the Krawtchouk model to be described below, the reader is encouraged to look at \cite{8} where it was originally introduced in the PST context.) Note that given \eqref{39}, $|\bar{P}'_{N+1}(\bar{x}_s)|=s!(N-s)!$. In view of formula \eqref{24}, the weights will be given by the binomial distribution
\begin{align} \label{40}
 \bar{w}_s = \frac{N!}{s!(N-s)!}\left(\frac{1}{2}\right)^{N}
\end{align}
where the normalization constant $N!(\tfrac{1}{2})^N$ has been determined from $\sum_{s=0}^N \bar{w}_s = 1$. The polynomials orthogonal with respect to those weights are known to be the normalized symmetric Krawtchouk polynomials
\begin{align}\label{41}
 &\hat{K}_n(s) = (-1)^n \sqrt{\binom{N}{n}} \pFq{2}{1}{-n,-s}{-N}{2} \quad \notag\\&n,s=0,\dots,N
\end{align}
where $\binom{l}{k}= l!/k!(l-k)!$ is the binomial coefficient and $_2F_1$ is the hypergeometric series
\begin{align} \label{42}
 \pFq{2}{1}{a,b}{c}{z} = \sum_{k=0}^\infty \frac{(a)_k(b)_k}{(c)_k}\frac{z^k}{k!}
\end{align}
with $(a)_k$ the Pochhammer symbol defined by
\begin{align}
 &(a)_0 =1 \quad (a)_k = a(a+1)\dots(a+k-1) \notag\\ &k =1,2,\dots 
\end{align}
Note that $\pFq{2}{1}{a,b}{c}{z}$ terminates when $a$ or $b$ is a negative integer. One has 
\begin{align} \label{44}
 \sum_s \frac{N!}{s!(N-s)!}\left(\frac{1}{2}\right)^N \hat{K}_m(s)\hat{K}_n(s) = \delta_{nm}.
\end{align}
Now it can be checked directly or by consulting Ref. \cite{21} that these polynomials $\hat{K}_n(s)$ satisfy the three-term recurrence relation
\begin{align} \label{45}
 (s-\frac{N}{2})\hat{K}_n(s) = a_{n+1}\hat{K}_{n+1}(s) +a_n\hat{K}_{n-1}(s)
\end{align}
with 
\begin{align}\label{46}
 a_n = \frac{1}{2}\sqrt{n(N-n+1)}.
\end{align}
With $\bar{\chi}_n(\bar{x}_s)$ identified as $\hat{K}_n(s)$, this implies that the entries of the matrix $\bar{J}$ are provided by the coefficients
\begin{align} \label{47}
 \bar{J}_n = a_n, \qquad \bar{B}_n =0.
\end{align}
It is immediate to check that these $\bar{J}_n$ and $\bar{B}_n$ verify \eqref{19} and hence define a mirror-symmetric $\bar{J}$.

Alternatively, one can also use quantum angular momentum theory to relate the spectrum \eqref{39} to the matrix $\bar{J}$ with elements given by \eqref{46} and \eqref{47}. Indeed observe that this $(N+1)$-dimensional matrix $\bar{J}$ coincides with that of the angular momentum operator $J_x$ in the standard basis $\ket{j,m} = \ket{\tfrac{N}{2},s-\tfrac{N}{2}}$, $s=0,\dots,N$, where the $z$-projection $J_z$ is diagonalized : $J_z\ket{j,m} = m\ket{j,m}$. Since all angular momentum components have the same spectrum, it follows that $\bar{J}$ has \eqref{39}, that is $(-\tfrac{N}{2},-\tfrac{N}{2}+1,\dots,\tfrac{N}{2}-1,\tfrac{N}{2})$ as eigenvalues. Now the eigenvectors of $J_x$ are obtained by transforming the vectors $\ket{j,m}$ under a rotation that takes the $z$-axis into the $x$-axis. Since elements of matrices irreducibly representing rotations are expressed in terms of Krawtchouk polynomials \cite{22}, their occurence in our problem is thus understood.

\subsection{The nearest-neighbour case}

When $M=1$ and the Hamiltonian $H^{(1)}$ is defined from \eqref{20} by taking 
\begin{align} \label{48}
 Q_1(\bar{J}) = \beta \bar{J}
\end{align}
with $\beta$ an arbitrary positive real number, it is immediate to recover the well known fact that the Krawtchouk model with nearest neighbour couplings admits PST. Indeed, condition \eqref{35} reads then
\begin{align} \label{49}
 T\beta(s-\tfrac{N}{2}) = -\phi + \pi (N+s+2L_s).
\end{align}
This shows that the integer numbers $L_s$ must necessarily depend linearly on $s$ and thus take the form $L_s = ls+m$ with $l$ and $m$ integers. With $\phi$ appropriately chosen to take care of the constant terms, \eqref{49} reveals that PST will be achieved at times $T$ given by
\begin{align} \label{50}
 T = \frac{\pi}{\beta}(2l+1) \qquad l=0,1,\dots
\end{align}
Hence the minimal time for PST in the NN model is $T=\tfrac{\pi}{\beta}$.

Let us here remark that $T$ (or equivalently $\beta$), the time for PST, and the integer $N$ that determines the length of the chain, are treated as two independent parameters. These two quantities are tied together however within the expression for the couplings : $J_n = \tfrac{\beta}{2}\sqrt{n(N-n+1)}$. The time $T$ can thus be kept fixed for different $N$ at the expense of changing the Hamiltonian. Note in this connection that the middle couplings grow with $N$. Therefore, in order to keep the couplings relatively small, we can either take $T$ proportional to $N$ or, keeping $T$ and $N$ independent, consider chains where $N$ is not too large. The latter view is typically the one adopted with the idea that spin chains are devices aimed at quantum transport over short distances \cite{1}.

\subsection{The next-to-nearest neighbour extension}

We are now ready to provide an analytic NNN extension of the Krawtchouk chain and to determine the specifications for which this spin chain will possess PST.

Let $M=2$ and take $Q_2(\bar{J})$ to be 
\begin{align} \label{51}
 Q_2(\bar{J}) = \alpha \bar{J}^2 + \beta \bar{J}
\end{align}
with $\alpha$ an arbitrary positive real number and $\beta$ a nonnegative number. According to \eqref{28}, $H^{(2)}$ has the following action on the one-excitation states $\ket{n}$: 
\begin{align} \label{52}
 H^{(2)}\ket{n} = &\alpha a_{n+1}a_{n+2}\ket{n+2} + \beta a_{n+1}\ket{n+1}\notag \\&+ \alpha (a_n^2 + a_{n+1}^2)\ket{n} + \beta a_n\ket{n-1} \notag\\&+ \alpha a_{n-1}a_n\ket{n-2}.
\end{align}
This defines a spin chain with NNN interactions that extends the Krawtchouk chain with NN links. The couplings and magnetic field strengths of the Hamiltonian $H^{(2)}$ can be read off from \eqref{52} by comparing with \eqref{4}; one finds
\begin{subequations}\label{53}
\begin{align}
 J_n^{(1)} &= J_n = \beta a_n \label{53a}\\
 J_n^{(2)} &= \alpha a_{n-1}a_n. \label{53b}
\end{align}
\end{subequations}
Note that the NN Hamiltonian $H^{(1)}$ is recovered when $\alpha=0$, observe moreover that $H^{(2)}$ has magnetic field strengths $B_n$ given by
\begin{align} \label{54}
 B_n = \alpha(a_n^2+a_{n+1}^2)
\end{align}
in contrast to $H^{(1)}$ where $B_n=0$.

It is known that quantum walks generated by spin chains and classical birth and death processes are intimately connected \cite{23}. Let us mention in this respect that our construction of the analytic Hamiltonian $H^{(2)}$ has similarities with the generalization of the Ehrenfest urn model developed in Ref. \cite{24}. This last paper offers an exact solution of a Markov process that involves nearest and next-to-nearest neighbours. Although the analysis in Ref. \cite{24} is framed in terms of matrix orthogonal polynomials, the pentadiagonal one-step transition probability matrix is in fact obtained, up to a constant term, as a quadratic expression with fixed coefficients in the Jacobi matrix of unnormalized Krawtchouk polynomials.

Furthermore, the method proposed in \cite{MS} to produce multi-particle entangled states of ions in an ion trap is based on the use of $J_x^2$ as Hamiltonian. From our discussion at the end of subsection 3.2, we see that this relates to the case $\beta=0$. The study in \cite{MS} thus has connections to our considerations.

The analysis of this NNN model will lead us to results regarding PST that can be summarized as follows.
% \begin{center}
%\begin{minipage}{0.7\textwidth}
For $\beta>0$, $H^{(2)}$ will generate PST if $\tfrac{\alpha}{\beta}$ is a rational number, that is, if $\tfrac{\alpha}{\beta} = \tfrac{p}{q}$ where $p$ and $q$ are two co-prime integers. Moreover if $p$ is odd, $q$ and $N$ will need to be of the same parity. PST will then be observed at $T = \tfrac{\pi q}{\beta}$. For PST to happen in the NNN chain at the same time $\tfrac{\pi}{\beta}$ as in the NN model, we must have $\tfrac{\alpha}{\beta}=p$, $p\in\mathbb{N}$, with the odd $p$ only admissible when $N$ is odd. For $\beta=0$, there will be PST for even $N$ with minimal time $T=\frac{\pi}{\alpha}$, but no PST for odd $N$.
%\end{minipage}
%    \end{center}

Let us now explain how this is found by examining if there are values of the parameters $\alpha$ and $\beta$ for which $H^{(2)}$ will generate PST. The general condition \eqref{35} for PST becomes 
\begin{align} \label{55}
 T[\alpha(s-\tfrac{N}{2})^2 + \beta(s-\tfrac{N}{2})] = -\phi +\pi N + \pi s + 2\pi L_s
\end{align}
when $Q_2$ is given by \eqref{51}. Since the lhs of \eqref{55} is a quadratic polynomial in $s$, we need to take $L_s$ to be also quadratic. Let
\begin{align} \label{56}
 L_s = \xi s^2 + \eta s + \zeta .
\end{align}
It is easy to show that $L_s$ will be an integer for all $s=0,1,\dots,N$ when $\zeta$ is an integer and $\xi$ and $\eta$ are simultaneously integers or half-integers. Using \eqref{56} and equating the coefficients of the various powers of $s$ in \eqref{55} leads to the relations
\begin{subequations}
\begin{align}
 \alpha T &= 2\pi \xi \label{57a} \\
 2\pi\eta + \pi &= -\alpha TN+\beta T \label{57b} \\
 \alpha T \tfrac{N^2}{4} - \beta \tfrac{N}{2} &= -\phi + \pi N + 2\pi\zeta. \label{57c}
\end{align}
\end{subequations}
The last equation \eqref{57c} does not imply any restrictions as it merely provides a relation between the coefficient $\zeta$ and the phase $\phi$. From \eqref{57a} and \eqref{57b} we see that
\begin{align}\label{58}
 \frac{\beta}{\alpha} = \frac{2\eta+1}{2\xi} + N
\end{align}
which indicates that the ratio $\tfrac{\beta}{\alpha}$ should be a rational number (depending on the choice of $\xi$ and $\eta$). From here onwards we will analyse the cases $\beta>0$ and $\beta=0$ separately, starting with $\beta>0$.

The time $T$ for PST is given by 
\begin{align} \label{59}
 T = \frac{2\pi}{\beta}[N\xi+\eta+\tfrac{1}{2}].
\end{align}
Because of the limitations on the values of $\xi$ and $\eta$, it is clear that only two cases are possible : 
\begin{align*}
 (i&) \quad T = \frac{2\pi j}{\beta} \\
 (ii&)\quad T = \frac{\pi}{\beta}(2j+1)
\end{align*}
with $j$ some integer. Indeed, when both $\xi$ and $\eta$ are integers only case $(ii)$ is realized; when $\xi$ and $\eta$ are half-integers case $(i)$ occurs when $N$ is even and case $(ii)$  happens when $N$ is odd. The smallest time $T$ for which PST can be obtained arises from case $(ii)$ when $j=0$.  This yields the same minimal PST time as for the NN model which is $T = \tfrac{\pi}{\beta}$. Eq. \eqref{59} shows that this implies
\begin{align} \label{60}
 \eta = -N\xi.
\end{align}
When $N$ is odd, \eqref{60} shows that $\xi$ and $\eta$ can both be chosen to be integers or half-integers. When $N$ is even, then necessarily $\xi$ and $\eta$ must be integers. Whether $N$ is even or odd, we always have owing to \eqref{60}
\begin{align} \label{61}
 \frac{\alpha}{\beta} = 2\xi.
\end{align}
We thus arrive at the following results. In the NNN chains with parameters $\alpha$ and $\beta$, PST is achieved at the same time $T = \tfrac{\pi}{\beta}$ as in the NN chain (with parameter $\beta$) if $\alpha$ is in the following relation with $\beta$ : 
\begin{align*}
 (i&) \quad \frac{\alpha}{\beta} = 1,2,3,\dots \qquad \text{for $N$ odd} \\
 (ii&)\quad \frac{\alpha}{\beta} = 2,4,6,\dots \qquad \text{for $N$ even.} 
\end{align*}
There are other circumstances when PST can be achieved of course. Recall that \eqref{48} simply required $\tfrac{\alpha}{\beta}$ to be rational. In general, if $\tfrac{\alpha}{\beta} = \tfrac{p}{q}$ where $p$ and $q$ are co-prime integers, take
\begin{align} \label{62}
 \xi = \tfrac{p}{2} \quad \text{and} \quad \eta = \tfrac{1}{2}[q-Np-1],
\end{align}
so that $N\xi +\eta+\tfrac{1}{2} = \tfrac{q}{2}$ and \eqref{58} is satisfied. If $p$ is even, $q$ must be odd; $\xi$ and $\eta$ are then both integer. If $p$ is odd, $\xi$ is half-integer; $\eta$ will also be half-integer either for $q$ odd and $N$ odd or, for $q$ even and $N$ even. In all cases, $T = \tfrac{\pi q}{\beta}$.

Let us now look at $\beta=0$ in which case \eqref{57a} and \eqref{57b} yield
\begin{align}\label{335}
2 \eta+1=-2\xi N                
\end{align}
and
\begin{align}\label{336}
 T = \frac{2\pi\xi}{\alpha}.
\end{align}
Recall that $\xi$ and $\eta$ are either both integers or both half-integers. In the former case, the left hand side of \eqref{335} is odd while its right hand side is even. This contradiction rules out the option $\xi,\eta$ integers. When $\xi$ and $\eta$ are both of the form $\xi=u/2$ and $\eta=v/2$ with $u$ and $v$ odd integers, we see that \eqref{335} becomes $u+1=-vN$ which can only be fulfilled if $N$ is even. The minimal PST time in this instance is $T=\tfrac{\pi}{\alpha}$ which is obtained for $\xi=1/2$. 

Note that perfect state transfer will not occur if $\tfrac{\alpha}{\beta}$ is not rational. This is to be contrasted with the fact that for the NN-Hamiltonian ($\alpha=0$), PST is always observed. The possibilities of almost perfect state transfer (APST) (see \cite{25,26}) when $\tfrac{\alpha}{\beta}$ is irrational remain to be investigated.

\section{Fractional revival} \label{sec:FR}
\subsection{General considerations}
Fractional revival (FR) can also be observed in certain spin chains. In fact, PST can be viewed as a special case of FR, a phenomenon that sees the time evolution of a wave-packet generate periodically a number of "smaller reproductions" of the initial state at specific locations. In the PST situation, only one unscathed reproduction is oberved at given times at the ends of the chain. We shall examine in this section the possibility of FR at two sites in the NNN spin chains on which we have focused so far. To avoid confusion with PST, we shall denote by $\tau$ the FR time. FR at the sites $0$ and $N$ will be realized at time $\tau$ if
\begin{align} \label{63}
 e^{-iH\tau}\ket{0} = \mu\ket{0} + \nu\ket{N}
\end{align}
where the complex amplitudes $\mu$ and $\nu$ are subjected to the condition
\begin{align} \label{64}
 |\mu|^2 + |\nu|^2 =1. 
\end{align}
Relation \eqref{63} indicates that the initial state with one spin up at site $0$ evolves after time $\tau$ into a state which is described by a linear combination of two state vectors associated respectively to a spin up at site $0$ and another spin up at site $N$. PST corresponds to $\mu=0$ ($|\nu|=1 $) and when $\nu=0$ ($|\mu|=1 $), we have a perfect return. Replacing the entries $0$ and $1$ by $\uparrow$ and $\downarrow$ in the vectors \eqref{3} (and forgetting the transposition), it is readily recognized that when $\mu=\nu=1/\sqrt{2}$, the state into which $\ket{0}$ evolves, namely
\begin{align*}
 \frac{\ket{0}+\ket{N}}{\sqrt{2}} = \frac{\ket{\uparrow\downarrow\dots\downarrow}+\ket{\downarrow\dots\downarrow\uparrow}}{\sqrt{2}}
\end{align*}
is maximally entangled.

The conditions for fractional revival at two sites in NN spin chains of XX type have been thoroughly analyzed in \cite{7}. In general, the one-excitation Hamiltonian is only required to be an isospectral deformation of a mirror-symmetric Jacobi matrix. Here we wish to study the occurence of FR in spin chains with non-nearest neighbour interactions that belong to the special class introduced in Section \ref{sec:prelim} and that have couplings built from polynomials in a mirror-symmetric matrix $\bar{J}$ (see \eqref{28}). In fact, we want to concentrate in the end on the NNN model with $H^{(2)}\ket{n} = (\alpha\bar{J}^2 + \beta\bar{J})\ket{n}$ and $\bar{J}$ defined by \eqref{46} and \eqref{47}, in order to determine for what values of the parameters $\alpha$ and $\beta$ will FR occur.

Under the assumption that $\bar{J}$ is mirror-symmetric, the associated polynomials $\bar{\chi}_n(x_s)$ $n=0,1,\dots,N$ satisfy \eqref{25}, that is $\bar{\chi}_N(x_s)=(-1)^{N+s}$. Using the expansion \eqref{12} as in Section \ref{sec:prelim} and $H^{(M)}\ket{n} = Q_M(\bar{J})\ket{n}$ we see that \eqref{63} amounts to 
\begin{align} \label{65}
 e^{-i\tau Q_M(x_s)} = e^{i\phi} (\mu'+\nu'\bar{\chi}_N(x_s))
\end{align}
where $\mu= e^{i\phi}\mu'$ and $\nu= e^{i\phi}\nu'$. Note that $\mu'$ can be taken real once a global phase term $e^{i\phi}$ has been factored. Taking the modulus of both sides of \eqref{65} and using $\bar{\chi}_N^2(x_s) =1$, we find that
\begin{align} \label{66}
 \text{Re}(\mu'\nu')=0.
\end{align}
Since $\mu'$ is taken to be real, $\nu'$ must be an imaginary number. In view of \eqref{64}, we shall write
\begin{align} \label{67}
 \mu' = \cos\theta \quad \nu'=i\sin\theta
\end{align}
and given that $\bar{\chi}_N(x_s)=(-1)^{N+s}$, \eqref{65} will read
\begin{align} \label{68}
 e^{-i\tau Q_M(x_s)} = e^{i\phi} (\cos\theta +i(-1)^{N+s}\sin\theta).
\end{align}
In this parametrization, up to integer multiples of $\pi$, $\theta=\tfrac{\pi}{2}$ corresponds to PST and $\theta=0$ implies a perfect return at time $\tau$.

Let us now focus our attention on the model discussed before with $x_s = s-\tfrac{N}{2}$, $s=0,\dots,N$.

\subsection{The nearest-neighbour Krawtchouk chain}
Consider first the case $M=1$ with $Q_1(\bar{J}) = \beta \bar{J}$. Condition \eqref{68} becomes 
\begin{align} \label{69}
 e^{-i\tau \beta(s-\tfrac{N}{2})} = e^{i\phi} (\cos\theta + i(-1)^{N+s}\sin\theta).
\end{align}
This equation splits into the following two relations according to the parity of $s$: 
\begin{subequations}
\begin{align}
 \beta\tau(2s-\tfrac{N}{2}) &= -\phi - (-1)^N\theta + 2\pi L_s^{(0)} \label{70a} \\
 \beta\tau(2s+1-\tfrac{N}{2}) &= -\phi + (-1)^N\theta + 2\pi L_s^{(1)} \label{70b}
\end{align}
\end{subequations}
where $L_s^{(i)}$, $i=0,1,$  are two a priori independent sequences of integers that must be of the form 
\begin{align} \label{71}
 L_s^{(i)} = \gamma_i s + \delta_i \qquad i=0,1
\end{align}
with $\gamma_i$ and $\delta_i$ integers. It follows from  \eqref{70a}  and  \eqref{70b} that $\gamma_0=\gamma_1=1,2\dots$ and that 
\begin{align}\label{72}
 \tau = \pi\frac{\gamma_0}{\beta}.
\end{align}
  Moreover, apart from a relation determining the phase $\phi$ in terms of the parameters, one finds that $\theta$ is given by
\begin{align} \label{73}
 \theta = (-1)^N \left[ \frac{\gamma_0}{2} + (\delta_0-\delta_1) \right]\pi.
\end{align}
Therefore, up to sign and integer multiples of $\pi$, $\theta$ can only take two distinct values, namely 0 and $\tfrac{\pi}{2}$. This means that only PST and perfect return are possible. We thus reach the conclusion that FR at two sites cannot happen in the Krawtchouk NN model.

\subsection{The next-to-nearest-neighbour Krawtchouk chain}
We now set $M=2$ and take $Q_2(\bar{J}) = \alpha\bar{J}^2 +\beta\bar{J}$. In this case, the FR condition \eqref{68} yields the two relations: 
\begin{subequations} \label{74}
 \begin{align}
  \alpha\tau(2s-\tfrac{N}{2})^2 &+ \beta\tau(2s-\tfrac{N}{2}) =\notag\\ &-\phi -(-1)^N\theta +2\pi M_s^{(0)} \label{74a} \\[1em]
  \alpha\tau(2s+1-\tfrac{N}{2})^2 &+ \beta\tau(2s+1-\tfrac{N}{2}) =\notag\\ &-\phi +(-1)^N\theta +2\pi M_s^{(1)} \label{74b} 
 \end{align}
\end{subequations}
where anew, $M_s^{(0)}$ and $M_s^{(1)}$ are sequences of integers. In this case for the two sides of eqs.\eqref{74} to be compatible, we must take quadratic expressions
\begin{align} \label{75}
 M_s^{(i)} = \xi_i s^2+\eta_i s +\zeta_i \quad i=0,1
\end{align}
where for each $i$, independently, $\xi_i$ and $\eta_i$ can in general be simultaneously integer or half-integer while $\zeta_i$ is integer.

Once \eqref{75} is used, equating the coefficients of the powers of $s$ in \eqref{74} gives a system of 6 equations. It is easy to see that they amount to
\begin{align}
 \xi_0 &= \xi_1 \qquad\quad \eta_1-\eta_0 = \xi_0 \label{76}\\
 \xi_0 &= \frac{2\alpha\tau}{\pi} \qquad \eta_0 = (\beta-\alpha N)\frac{\tau}{\pi} \label{77}\\
 \zeta_1 - \zeta_0 &= \frac{1}{2\pi}\left[ (\alpha+\beta)\tau -\alpha N \tau -2(-1)^N\theta \right] \label{78}
\end{align}
with the sixth equation fixing the phase $\phi$ in terms of the parameters $\alpha,\beta,\tau,N,\theta,\zeta_0$ and $\zeta_1$.

The relations \eqref{76} imply that all the parameters $\xi_i,\eta_i,\zeta_i,i=0,1$ are integers. Indeed, assume that $\xi_0$ and $\eta_0$ are half-integers; $\xi_1=\xi_0$ and since $\xi_1$ is hence half-integer, $\eta_1$ must also be half-integer. Since the difference $\eta_1-\eta_0$ of two half-integers is an integer, we have a contradiction with $\eta_1-\eta_0=\xi_0$. All parameters can therefore only be integer.

From eqs.\eqref{77}, one obtains when $\beta>0$, the following expression for the FR time $\tau$ and the relation \eqref{80} between $\alpha$ and $\beta$ :
\begin{align}
 \tau = \frac{\pi}{\beta}(\eta_0 + \tfrac{N}{2}\xi_0) \label{79}\\
 \frac{\alpha}{\beta} = \frac{\xi_0}{2\eta_0+N\xi_0} \label{80}.
\end{align}
When $\beta=0$, one gets 
\begin{align}
\tau=\frac{\pi}{2 \alpha} \xi_0 \\
\eta_0=-\frac{N}{2} \xi_0. \label{420}
\end{align}
From \eqref{78} and using \eqref{67}, one finds for $\theta$ in both cases :
\begin{align} \label{81}
 \theta = (-1)^N \pi [\tfrac{\xi_0}{4} + \tfrac{\eta_0}{2} + \zeta_0-\zeta_1].
\end{align}

We can now draw some conclusions from these formulas on the occurrence of FR in the NNN model. 

%\begin{center}
%\begin{minipage}{0.7\textwidth}
For $\beta>0$: (i) First, we find the same constraint on $\alpha$ and $\beta$ for the presence of FR as for PST, namely that $\tfrac{\alpha}{\beta}$ is a rational number; 
(ii) Second, we observe that the fractional revival times will be integer multiples of $\tfrac{\pi}{2\beta}$ which is half the minimal PST time; 
(iii) Third, and most importantly, we note that FR at two sites is genuinely possible. Indeed, up to signs and the addition of integer multiples of $\pi$, we see from \eqref{81} that $\theta$ can take the values $0,\tfrac{\pi}{4},\tfrac{\pi}{2}$. We already observed that $\theta=0$ and $\theta=\tfrac{\pi}{2}$ correspond to perfect return and PST respectively. All possibilities that yield a  $\theta$ equivalent to $\tfrac{\pi}{4}$ will give rise however to balanced fractional revival where the amplitudes for finding a spin up at the sites 0 and $N$ are both equal in magnitude to $1/\sqrt{2}$. For $\beta=0$, since $\xi_0$ and $\eta_0$ are integers, eq.\eqref{420} requires $N$ to be even. Therefore : (i) when $N$ is odd no FR occurs; (ii) when $N$ is even $\theta$ can be equal to $\tfrac{\pi}{4}$ modulo multiples of $\pi$ and FR is possible, in this case the minimal FR time is $\tfrac{\pi}{2\alpha}$ with PST happening at time $\tfrac{\pi}{\alpha}$.  
%\end{minipage}
%    \end{center}

The realization of a specific scenario will depend on the characteristics of the NNN model as they are determined by the integer parameters $\xi_i,\eta_i$ and $\zeta_i$. From the preceding discussion, we note that FR at sites 0 and $N$ will occur only if $\xi_0$ is odd. That is, in view of \eqref{80}, FR will be seen only in the NNN spin chains where $\tfrac{\alpha}{\beta}= \tfrac{p}{q}$ with $p$ and $q$ co-prime and $p$ odd. To achieve such a value of $\tfrac{\alpha}{\beta}$ we may then set
\begin{align} \label{82}
 \xi_0 = p \qquad \eta_0 = \tfrac{1}{2}(q-Np)                                                                                                                                                                                                                                                                                                                                                                                                                                                                                                                                                                                                                                                                                                                                                                                                                                                                                                                                                                            \end{align}
 in order to satisfy \eqref{80}. Since $p$ is odd, $\eta_0$ will be integer as required either when $q$ and $N$ are both even or when $q$ and $N$ are both odd. Irrespective, we see from \eqref{79} that the FR time $\tau$ is given by $\tau = \tfrac{\pi q}{2\beta}$. One observes that this is consistent with the discussion of the conditions for PST; in that case the specifications involved the parameters $\xi$ and $\eta$ that could be both integers or both half-integers. For $p$ even and $q$ odd, we note that while PST is possible, there will be no FR. The other PST cases will exhibit FR in addition. Comparing \eqref{62} and \eqref{82}, we see that
\begin{align}
 \xi = \frac{\xi_0}{2} \qquad \eta = \eta_0 - \tfrac{1}{2}.
\end{align}
Since $\xi_0$ and $\eta_0$ are integers and $\xi_0$ is odd, this makes $\xi$ and $\eta$ half-integers in keeping with what was found earlier. Observe also that the parity considerations on $N$ match. It is then seen that the PST time $T$ is double the FR time $\tau$ : $T=2\tau$.

This last point can also be understood as follows. From \eqref{80}, it is clear that we can multiply both parameters $\xi_0$ and $\eta_0$ by the same integer $k$ : $\xi_0\to k\xi_0, \eta_0\to k\eta_0$ without changing the ratio $\tfrac{\alpha}{\beta}$, that is without changing the Hamiltonian essentially. This transformation however will have the effect of scaling $\tau$ as seen from \eqref{79} : $\tau \to k\tau$. Consider now what happens to $\theta$ under that transformation when $k=2$:
\begin{align}
 \theta \to \bar{\theta} = (-1)^{N+1}\pi(\tfrac{\xi_0}{2}+\eta_0+\zeta_0-\zeta_1). \label{84}
\end{align}
 Assume that we are in a situation of FR and thus that $\xi_0$ is odd. It follows that $\cos\bar{\theta}=0$. Such a $\bar{\theta}$ leads to PST since the only non-zero spin up amplitude is then at site $N$. 

We thus have the following scenario. If fractional revival occurs at time $t=\tau$, then at $t=T=2\tau$ we shall observe PST, at $t=3\tau$ fractional revival will be seen again and finally at $t=2T=4\tau$ a perfect return will happen. This cycle will then repeat itself with period $4\tau$. 

In summary, we have made in this section the following interesting observations. We have found that while FR does not occur in the NN Krawtchouk spin chain, the presence of additional NNN interactions allow this phenomenon to take place. Furthermore, the only form of FR at sites 0 and $N$ that can be realized is of the balanced type which corresponds to the generation of a maximally entangled state. Finally, when NN interactions are kept, that is when $\beta>0$, for FR to occur the parameters $\alpha$ and $\beta$ of $H^{(2)}$ must satisfy $\tfrac{\alpha}{\beta}=\tfrac{p}{q}$ with $p$ and $q$ coprime integers, $p$ odd and $q$ and $N$ both odd or even; FR will then happen at time $\tau = \tfrac{\pi q}{2\beta}$.

\section{Conclusion}\label{sec:conclusion}

We succeeded with our goal of providing an analytic spin chain with couplings beyond the nearest neighbours where an exact description of PST and FR could be given. We focused on models where the higher order interactions are related in a polynomial fashion to those of the nearest neighbours and we concentrated on the one model in that class which extended with NNN links, the simplest non-uniform and well-studied XX spin chain with NN couplings based on angular momentum theory or the recurrence coefficients of the Krawtchouk polynomials. This extended model involves two parameters $\alpha$ and $\beta$ that tune the intensity of the NNN and NN couplings respectively. The case $\alpha=0$ corresponds to the absence of second order interactions. One recalls that PST is then known to occur first at time $T=\tfrac{\pi}{\beta}$ and we indicated that no FR is predicted by this most simple model. 

When $\alpha\neq 0$ and $\beta\neq0$, the conditions for PST and FR can be summarized easily. In order for PST to occur, one must have $\tfrac{\alpha}{\beta} = \tfrac{p}{q}$ where $p$ and $q$ are co-prime integers. Furthermore, if FR is to happen, $p$ must be odd and in that case $N$ must be of the same parity as $q$.

%We explored in addition how this NNN model could inform the quantum transport in arrays of evanescently coupled waveguides where couplings beyond first order, while generally small, will necessarily be present. The examination of the analytic spin chain with NNN couplings showed that it is well adapted to this optical waveguide context and allowed to discuss experimental observations. Moreover, it led to the remarkable conclusion that in principle FR should be occur in photonic lattices and should provide a mechanism for the generation of maximally entangled states. This is in contrast with what the simplest NN Krawtchouk model predicts in this regard. We thus feel that it would be of great interest to look for FR in optical arrays. There are no doubt various technical difficulties in the way of an experimental confirmation; one may hope nevertheless that it can be accomplished as was done for PST. On the theoretical side this motivates extending the present study in many directions. 

It would be useful to perform various simulations in order to probe the robustness of the PST and FR phenomena with respect to deviations from the specifications of the analytic model. In that regard it would be worthwhile to explore in details the conditions for almost perfect transfer and fractional revival. Finally, it would be relevant to construct other NNN analytic models so as to offer experimentalists a wider array of possibilities for tuning and concrete realizations. We plan to pursue some of these questions.

\section*{Acknowledgments}
LV has much appreciated the hospitality of the Department of Mathematical Sciences of the University of Copenhagen while this work was being completed. AZ was a visiting researcher of the Centre de Recherches Math\'ematiques (CRM) when this project was developed. The authors are grateful to Maxim Derevyagin, Jean-Michel Lemay and Anders S\o rensen for stimulating discussions. MC acknowledges financial support from the European Research Council (ERC Grant Agreement no 337603), the Danish Council for Independent Research (Sapere Aude) and the
Swiss National Science Foundation (project no PP00P2\_150734). The research of LV is supported in part by a research grant from the Natural Sciences and Engineering Research Council (NSERC) of Canada.

\bibliography{NNN.bib}
\bibliographystyle{plain}

\end{document}